# Quantum Monte Carlo Investigation of the H-transfer Reaction of Criegee Intermediate CH$_3$CHOO: A Benchmark Calculation

Zhiping Wang[1,*] and Yuxiang Bu[2]

[1] *School of Physics, Shandong University, Jinan, 250100, People's Republic of China*
[2] *School of Chemistry and Chemical Engineering, Shandong University, Jinan 250100, People's Republic of China*

**ABSTRACT**

We perform the fixed-node diffusion Monte Carlo (FN DMC) calculations to determine the barrier height and reaction energy of a critical reaction, the H-transfer isomerization from syn-CH$_3$CHOO to vinyl hydroperoxide. The FN DMC barrier height is found to be 16.60±0.35 kcal/mol which agrees well with the experimental measurement within a few tenths of kcal, justifying the reliability of the FN DMC method for predicting barrier height of the rapid unimolecular reaction of Criegee intermediates. By comparing the predictions from the CCSD(T), G3 (MCG3), DFT and MP2 methods with respect to the FN DMC results and available experiment measurement, we found that while using Dunning's correlation consistent basis set aug-cc-pVTZ, the CCSD(T) barrier heights agree with the FN DMC counterpart within statistical errors, and is within a closer agreement with experiment and FN DMC prediction than the G3(MCG3) models. Barrier heights predicted from the relatively more economic DFT methods are within a few tenths' kcal of the FN DMC prediction. MP2 method severely underestimates the barrier height. FN DMC prediction for the reaction energy is -17.25±0.31 kcal/mol, setting an upper limit for the reaction energies predicted by the post Hartree-Fock methods and a lower limit for the DFT reaction energies. We provide FN DMC input for clarifying the energetic uncertainties in the critical H-transfer isomerization from syn-CH$_3$CHOO to vinyl hydroperoxide reaction. The quantitatively close agreements between the FN DMC barrier height and experimental measurement, and between the predictions from the FN DMC and G3 model for the reaction energy provide a theoretical basis for resolving the uncertainty in the energetics of such a type of important reactions.

# I. INTRODUCTION

Criegee intermediates (CIs) has drawn considerable attention because of its important role in atmosphere chemistry and its elusive characteristic.[1–5] They are the main intermediates during the ozonolysis of organic molecules. Since the preceding ozonolysis is highly exothermic, large portion of these intermediates CIs formed in the gas phase are initially hot molecules usually undergo rapid unimolecular reactions,[6] while a small portion of them will be thermally stabilized and then react with surrounding water, $SO_2$, $NO_2$, and volatile organic compounds in the atmosphere[4,7–10].

Among these rapid unimolecular reactions, the H-transfer channel is considered the most favorite reaction because of its low energy barrier, and it leads to vinly hydroperoxide (VHP) which will decompose to ·OH and vinoxy radical eventually. VHP is reported as another critical intermediate besides CIs during the ozonolysis of organic molecule in the atmosphere, as revealed by Drozd et al. and Donahue et al. through analysis of ·OH yields during the ozonolysis process.[11,12] As a very important source of the atmosphere detergent ·OH,[1–3] the H-transfer channel of CIs has been broadly studied in both theory and experiment, in aspects such as electronic structure, thermodynamic and reaction dynamic and energetics, *etc*.[13–16]

However, we found that the reaction energy and barrier height of the H-transfer isomerization predicted from different theoretical methods are very controversial. Take one of the prototypical and most stable Criegee intermediates syn-$CH_3CHOO$ as an example, the discrepancy in these energies is as high as 7 kcal/mol. The effective barrier height of the isomerization from syn-$CH_3CHOO$ to VHP was measured firstly by Liu *et al.*[17] and reported as 16 kcal/mol. They performed the state-selective excitation of cold syn-CH3CHOO with tunable infrared beam, in the CH stretching overtone region near 6000 cm$^{-1}$. By observing the action spectrum of syn-$CH_3CHOO$ they also identified the vibrational modes that coupled to the H-transfer reaction coordinates. In the same work, they also reported the theoretical barrier height and reaction energy determined at CCSD(T)/6-311+G(2d,p) on the stationary electronic structures at B3LYP/6-311+G(2d,p) level as 17.9 kcal/mol and -18.0 kcal/mol. They also reported the barrier height and reactions energies as 16.3 kcal/mol and 17.8 kcal/mol at the CCSD(T)//M06-2X/aug-cc-pVTZ level in a separate work.[18] No experimental data for the reaction energy has been seen in the literature so far. Kuwata et al. had been using a series of DFT methods as well as composite method, G3(MCG3) model to evaluate the energetics of this reaction, where the reported barrier height is ranging from 16.32 to 17.91 kcal/mol and reaction energy from -19.14 to -15.96 kcal/mol. The B3LYP and MP2 energies computed with a moderate basis set were also reported in an early work by Gutbrod *et.al.* Long *et. al.* had been using recently developed efficient composite wavefunction theory method to include beyond-CCSD(T) effects, and their W3X-L prediction for the barrier height is 17.01 kcal/mol.[19]

In the present work, we performed fixed-node diffusion Monte Carlo (FN DMC) calculations to determine the barrier height and reaction energy of the H-transfer reaction from syn-$CH_3CHOO$ to VHP, providing FN DMC input for clarifying the energetic uncertainties in this critical reaction. In the following sections we describe the computational methodology used in this study and discuss the findings and compared the present findings with those reported in literature.

## II. COMPUTATIONAL DETAILS

First, the stationary points of the reactant $CH_3CHOO$, product VHP and the transition state structure were obtained by structural optimization at B3LYP/6-311+(2d,p), with Gaussian 09[20], further validated by frequency analysis. Then fixed-node diffuse Monte Carlo[21–23] (FN DMC) calculations were performed on these structures to generate the corresponding FN DMC energy for each molecule. The FN DMC trial wave functions are products of antisymmetric and symmetric components. The former was chosen as a single determinant constructed from Kohn-Sham orbitals.

The latter was a 29-parameter Schmit-Monskowitz-Boys-Handy (SMBH) correlation function[24,25]. This expansion includes electron-electron and electron-nucleus terms and increases computational efficiency by reducing the variance of the local energy. The SMBH correlation function parameters were optimized using the linear optimization VMC algorithm of Toulouse and Umrigar. The GAMESS[26] *ab initio* package was used to generate the close-shell B3LYP orbitals using the triple-zeta energy-consistent pseudopotentials and basis sets of Burkatzki et al.[27] A FN DMC algorithm with small time-step errors was employed.[28] To maintain a high acceptance ratio, the electrons were moved one at a time. A simple branching algorithm[23] was used to duplicate walkers with large weights and to eliminate those with small weights. All QMC calculations were performed using the Zori code.[22] A FN DMC simulation ensemble of ~200000 walkers were chosen to ensure reduction of the population bias. Calculations were performed at 0.04, 0.02, 0.01 and 0.005 Hartree$^{-1}$ time-steps and were run until the stochastic error bars were below 0.0005 Hartree$^{-1}$. Weighted quadratic least-squares fits were used to extrapolate energy values to zero time-step.

In addition, we also carried out a series of structural optimizations with DFT and MP2 methods as well as CCSD(T) single point energy calculations for a comprehensive understanding of the performance of these methods with respect to the FN-DMC approach in characterizing the barrier height and reaction energy of the H-transfer reaction of the CI. The levels of theory used for structural optimizations are B3LYP/6-31G(d,p), B3LYP/6-31+G(d,p), B3LYP/6-311+G(2d,p), M06-2X/aug-cc-pVTZ, MP2/6-31G(d,p), MP2/6-31G+(d,p), MP2/6-311+G(2d,p) and MP2/aug-cc-pVTZ. Single point energies at the CCSD(T)/aug-cc-pVTZ level are obtained based on the structures optimized at the B3LYP/aug-cc-pVTZ and M06-2X/aug-cc-pVTZ levels, and also the CCSD(T)/6-311+G(2d,p) energies over the B3LYP/6-311+G(2d,p) structures. The absolute energies for each stationary point with various methods and the corresponding zero-point energy and ZPE scaling factor are listed in Table S2 in the Supporting information.

## III. RESULTS AND DISCUSSION

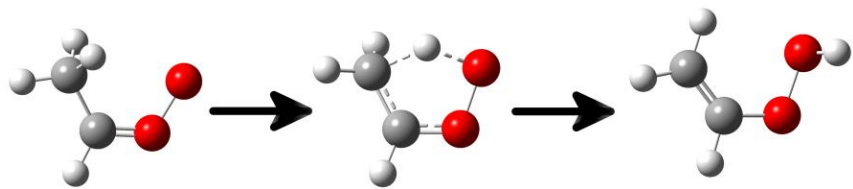

**Figure 1** Schematics of H-transfer reaction of $CH_3COO$, with stationary structures of syn-$CH_3CHOO$, VHP and the transition state. H, C and O atoms are in white, grey and red color.

Results of the FN DMC calculations are reported in Table 1 along with the findings with other theoretical methods as well as the effective barrier height from experimental measurement. The FN DMC barrier height of the isomerization from syn-$CH_3CHOO$ to VHP are 16.60±0.35, which are ZPE corrected at the B3LYP/6-311+G(2d,p) level. The FN DMC barrier height agree closely with the effective barrier height 16.0 kcal/mol, as measured by Liu et. al. using the action IR spectroscopy.[17]

For comparison, we also calculated the barrier heights at the CCSD(T)//B3LYP/aug-cc-pVTZ, CCSD(T)//B3LYP/6-311+G(2d,p) and CCSD(T)//M06-2X/aug-cc-pVTZ levels, which are 16.5 kcal/mol, 17.97 kcal/mol and 16.25 kcal/mol, respectively. We found that although the geometries of the stationary points optimized at the B3LYP/6-311+G(2d,p) and B3LYP/aug-cc-pVTZ levels are perfectly consistent with each other, the basis set difference brings in ~1.5 kcal/mol to the CCSD(T) barrier height, and that Dunning's correlation consistent basis set aug-cc-pVTZ should be used to better characterize the dynamical correlation of electrons in the CCSD(T) energy calculation. Using the Dunning's correlation consistent basis set aug-cc-pVTZ, the CCSD(T) prediction for the barrier height agrees with the FN DMC prediction within the statistical uncertainty.

**Table 1**. Zero-point energy corrected barrier height and reaction energy (kcal/mol) for the H-transfer reaction from syn-CH3CHOO to VHP, at various levels of theoretical approach. The experimental measurement of the effective barrier height is available for comparison.

|  | Barrier Heights | Reaction Energies |
| --- | --- | --- |
| FN DMC | 16.60±0.35 | -17.25±0.31 |
| Experiment[17] | 16.0 | -- |
| CCSD(T)//M06-2X/aug-cc-pVTZ | 16.25 (16.3[18]) | -18.86 (-17.8[18]) |
| CCSD(T)//B3LYP/aug-cc-pVTZ | 16.50 | -18.75 |
| CCSD(T)//B3LYP/6-311+G(2d,p) | 17.97 (17.9 [29]) | -17.6 (-18.0 [29]) |
| MCG3//QCISD/6-31G(d) [30] | 17.34 | -18.07 |
| MCG3//QCISD/MG3 [30] | 17.91 | -17.62 |
| W3X-L[19] | 17.01 | -- |
| B3LYP /aug-cc-pVTZ | 16.12 | -17.53 |
| B3LYP/6-311+G(2d,p) | 16.53 | -16.83 |
| B3LYP/6-31+G(d,p) | 16.74 (16.74 [30]) | -15.96 (-15.96 [30]) |
| B3LYP/6-31G(d,p) | 15.38 (14.8 [7]) | -16.27 (-16.1 [7]) |
| M06-2X/aug-cc-pVTZ | 15.18 | -21.95 |
| MPW1K/6-31+G(d,p) [30] | 16.32 | -19.14 |
| BB1K/6-31+G(d,p) [30] | 16.43 | -18.69 |
| MP2/6-31G(d,p) | 14.66 | -22.20 (22.0 [7]) |
| MP2/6-31G+(d,p) | 15.57 | -21.70 |
| MP2/6-311+G(2d,p) | 14.01 | -22.14 |
| MP2/aug-cc-pVTZ | 12.87 | -22.59 |

Kuwata *et. al.* had reported the G3 (MCG3) prediction for the barrier height at the MCG3//QCISD/6-31G(d) and MCG3//QCISD/MG3 level, which are 17.34 kcal/mol and 17.91 kcal/mol. Although the multicoefficient G3 (MCG3) method has been shown to predict highly accurate energetics for thermochemistry, thermochemical kinetics and hydrogen transfer barrier heights[31,32], for the hydrogen transfer reaction of Criegee intermediate, the CCSD(T) prediction is much closer to the FN DMC prediction and experimental measurement than the G3 models. In addition, Long *et. al.*[19] had been using recently developed efficient composite wavefunction theory method to include beyond-CCSD(T) effects, and their W3X-L prediction for the barrier height is 17.01 kcal/mol, does not outperform the CCSD(T) method in this case.

It is worth noting that B3LYP here has performance as good as the more contemporary MPW1K and BB1K methods. The latter two methods are reported by Truhlar and co-workers to produce accurate barrier heights for hydrogen transfer reactions involving radicals[33,34]. The B3LYP barrier height gradually converges to the effective experiment barrier height 16kcal/mol as the size of basis set increases. We also found that it's very important to include diffuse functions for the heavy atoms into the basis set. Without the diffuse functions to the non-H atoms, B3LYP will underestimate the barrier height by 1.4kcal/mol. For the MP2 method, increasing the size of basis set does not bring the barrier height to a closer agreement with the experiment or FN DMC prediction. Overall, MP2 predictions spread across a larger energy window, compared with all the other methods concerned here, and underestimate the barrier height by 0.5 to 3 kcal/mol depending on different basis set used, therefore, MP2 method should not be recommended for calculation of the energetics of the H transfer reaction of CI.

The FN DMC reaction energy of the H-transfer isomerization is -17.38±0.31 kcal/mol. Since experimental measurement data for this value is not currently available, we compared the predictions from other theoretical methods only to the FN DMC result. The CCSD(T) predictions are ~1.5kcal/mol lower than the former. The MCG3//QCISD/MG3 prediction -17.62 kcal/mol agrees with FN DMC within statistical errors. But using a different basis set MCG3//QCISD/6-31G(d), G3 model lower the reaction energy by 0.5 kcal/mol lower. The B3LYP results, depending on different basis sets used, are 0.4-1.3 kcal/mol higher than the FN DMC prediction, and outperforms the other DFT methods M062X, MPW1K and BB1K. The MP2 results again, lies the farthest from the FN DMC prediction, predicted the reaction energy to be 4.3 to 5.3 kcal/mol lower. Over all, The FN DMC prediction sets an upper limit for the reaction energies predicted by the post Hartree-Fock methods, and a lower limit for the DFT reaction energies. The good agreement between the predictions from FN DMC and G3 model provides a theoretical basis for resolving the uncertainty in the reaction energy.

## IV.   CONLUSION

We have performed fixed-node diffuse Monte Carlo calculations for the barrier height and reaction energy of a critical reaction, the H-transfer isomerization from syn-$CH_3CHOO$ to VHP. The FN DMC barrier height agrees with the experimental measurement within a few tenths of kcal, justifying the reliability of FN DMC method for predicting barrier height of this rapid unimolecular reaction of Criegee intermediate. By comparing the predictions from CCSD(T), G3 (MCG3), DFT and MP2methods to the FN DMC results and available experiment measurement, we found that CCSD(T) barrier heights agree with the FN DMC counterpart within statistical errors, while using Dunning's correlation consistent basis set aug-cc-pVTZ, and is within a closer agreement with experiment and FN DMC prediction than the G3(MCG3) models.  The discrepancy in the reaction energies predicted by the highly correlated methods CCSD(T), G3(MCG3) and FN DMC is ~1kcal/mol, with FN DMC prediction sets the upper limit.  The barrier height predicted from the relatively more economic DFT methods are within a few tenths' kcal of the FN DMC prediction, and with a basis set including the diffuse functions to the non-H atoms, B3LYP method has performance as good as the more contemporary MPW1K and BB1K methods. Over all, The FN

DMC prediction sets an upper limit for the reaction energies predicted by the post Hartree-Fock methods, and a lower limit for the DFT reaction energies. The predictions for both of the barrier height and reaction energy from the MP2 method lie the farthest from the FN DMC results, and should not be considered suitable for predicting the energetics of H-transfer reaction of CI. The good agreement between the predictions from FN DMC and G3 model provides a theoretical basis for resolving the uncertainty in the reaction energy. Experimental input for the reaction energy might be desirable for further confirmation and assessment of the performance of various theoretical models in characterizing the reaction energy of Criegee intermediate.

**AUTHOR INFORMATION**
**Corresponding Author**
*E-mail: wangzhp@mail.sdu.edu.cn

**REFERENCES**

[1] R. Criegee and G. Wenner, Justus Liebigs Ann. Chem. **564**, 9 (1949).

[2] M. Kanakidou, J.H. Seinfeld, S.N. Pandis, I. Barnes, F.J. Dentener, M.C. Facchini, R. Van Dingenen, B. Ervens, A. Nenes, C.J. Nielsen, E. Swietlicki, J.P. Putaud, Y. Balkanski, S. Fuzzi, J. Horth, G.K. Moortgat, R. Winterhalter, C.E.L. Myhre, K. Tsigaridis, E. Vignati, E.G. Stephanou, and J. Wilson, Atmos. Chem. Phys. **5**, 1053 (2005).

[3] M. Hallquist, J.C. Wenger, U. Baltensperger, Y. Rudich, D. Simpson, M. Claeys, J. Dommen, N.M. Donahue, C. George, A.H. Goldstein, J.F. Hamilton, H. Herrmann, T. Hoffmann, Y. Iinuma, M. Jang, M.E. Jenkin, J.L. Jimenez, A. Kiendler-Scharr, W. Maenhaut, G. McFiggans, T.F. Mentel, A. Monod, A.S.H. Prévôt, J.H. Seinfeld, J.D. Surratt, R. Szmigielski, and J. Wildt, Atmos. Chem. Phys. **9**, 5155 (2009).

[4] C.A. Taatjes, D.E. Shallcross, and C.J. Percival, Phys. Chem. Chem. Phys. **16**, 1704 (2014).

[5] D.L. Osborn and C.A. Taatjes, Int. Rev. Phys. Chem. **34**, 309 (2015).

[6] Z. Wang, Y.A. Dyakov, and Y. Bu, J. Phys. Chem. A **123**, 1085 (2019).

[7] R. Gutbrod, E. Kraka, R.N. Schindler, and D. Cremer, J. Am. Chem. Soc. **119**, 7330 (1997).

[8] Y.-P. Lee, J. Chem. Phys. **143**, 020901 (2015).

[9] L. Vereecken, D.R. Glowacki, and M.J. Pilling, Chem. Rev. **115**, 150406063223002 (2015).

[10] J.M. Anglada, J. González, and M. Torrent-Sucarrat, Phys. Chem. Chem. Phys. **13**, 13034 (2011).

[11] N.M. Donahue, G.T. Drozd, S.A. Epstein, A.A. Presto, and J.H. Kroll, in *Phys. Chem. Chem. Phys.* (2011), pp. 10848–10857.

[12] G.T. Drozd, J. Kroll, and N.M. Donahue, J. Phys. Chem. A **115**, 161 (2011).

[13] N. Sebbar, H. Bockhorn, and J.W. Bozzelli, Phys. Chem. Chem. Phys. **4**, 3691 (2002).

[14] J.M. Anglada, J.M. Boffil, S. Olivella, and A. Solé, J. Am. Chem. Soc. **118**, 4636 (1996).

[15] W.H. Richardson, J. Org. Chem. **60**, 4090 (1995).

[16] C.A. Taatjes, O. Welz, A.J. Eskola, J.D. Savee, A.M. Scheer, D.E. Shallcross, B. Rotavera, E.P.F. Lee, J.M. Dyke, D.K.W. Mok, D.L. Osborn, and C.J. Percival, Science (80-. ). **340**, 177 (2013).

[17] F. Liu, J.M. Beames, A.S. Petit, A.B. McCoy, and M.I. Lester, Science **345**, 1596 (2014).

[18] F. Liu, Y. Fang, M. Kumar, W.H. Thompson, and M.I. Lester, Phys. Chem. Chem. Phys. **17**, 20490 (2015).

[19] B. Long, J.L. Bao, and D.G. Truhlar, J. Am. Chem. Soc. **138**, 14409 (2016).

[20] 2009. Gaussian 09, Revision C.01, M. J. Frisch, G. W. Trucks, H. B. Schlegel, G. E. Scuseria, M. A. Robb, J. R. Cheeseman, G. Scalmani, V. Barone, B. Mennucci, G. A. Petersson, H. Nakatsuji, M. Caricato, X. Li, H. P. Hratchian, A. F. Izmaylov, J. Bloino, G. Zhe, (n.d.).

[21] B.L. Hammond, W.A. Lester, and P.J. Reynolds, \href{https://Doi.Org/10.1142/1170}{Monte Carlo Methods in Ab Initio Quantum Chemistry} (WORLD SCIENTIFIC, 1994).

[22] A. Aspuru-Guzik, R. Salomón-Ferrer, B. Austin, R. Perusquía-Flores, M. a Griffin, R. a Oliva, D. Skinner, D. Domin, and W. a Lester, J. Comput. Chem. **26**, 856 (2005).

[23] P.J. Reynolds, J. Chem. Phys. **77**, 5593 (1982).

[24] K.E. Schmidt and J.W. Moskowitz, J. Chem. Phys. **93**, 4172 (1990).

[25] S.F. Boys and N.C. Handy, Proc. R. Soc. A Math. Phys. Eng. Sci. **309**, 209 (1969).

[26] M.W. Schmidt, K.K. Baldridge, J.A. Boatz, S.T. Elbert, M.S. Gordon, J.H. Jensen, S. Koseki, N. Matsunaga, K.A. Nguyen, S. Su, T.L. Windus, M. Dupuis, and J.A. Montgomery, J. Comput. Chem. **14**, 1347 (1993).

[27] M. Burkatzki, C. Filippi, and M. Dolg, J. Chem. Phys. **126**, 234105 (2007).

[28] C.J. Umrigar, M.P. Nightingale, and K.J. Runge, J. Chem. Phys. **99**, 2865 (1993).

[29] F. Liu, J.M. Beames, A.M. Green, and M.I. Lester, J. Phys. Chem. A **118**, 2298 (2014).

[30] K.T. Kuwata, M.R. Hermes, M.J. Carlson, and C.K. Zogg, J. Phys. Chem. A **114**, 9192 (2010).

[31] B.J. Lynch and D.G. Truhlar, J. Phys. Chem. A **107**, 3898 (2003).

[32] B.J. Lynch and D.G. Truhlar, J. Phys. Chem. A **106**, 842 (2002).

[33] B.J. Lynch, P.L. Fast, M. Harris, and D.G. Truhlar, J. Phys. Chem. A **104**, 4811 (2000).

[34] Y. Zhao, B.J. Lynch, and D.G. Truhlar, J. Phys. Chem. A **108**, 2715 (2004).

**Supporting Information for: Quantum Monte Carlo Investigation of the H-transfer Reaction of Criegee Intermediate CH$_3$CHOO: A Benchmark Calculation**

Zhiping Wang [1,*] and Yuxiang Bu [2]

[1] *School of Physics, Shandong University, Jinan, 250100, People's Republic of China*
[2] *School of Chemistry and Chemical Engineering, Shandong University, Jinan 250100, People's Republic of China*

**Table S1.** Cartesian coordinates of syn-CH$_3$CHOO, vinyl hydroperoxide (VHP) and transition state used in the FN DMC calculations, which are obtained through geometry optimization at B3LYP/6-311+G(2d,p) level.

| Syn-CH$_3$CHOO | | | |
|---|---|---|---|
| C | 0.48145600 | 0.69842100 | -0.00001700 |
| O | -0.77542600 | 0.58826900 | -0.00000500 |
| H | 0.81396900 | 1.73129300 | 0.00011300 |
| C | 1.36474100 | -0.47268900 | -0.00000500 |
| H | 1.12996900 | -1.09900200 | -0.86837100 |
| H | 1.13049200 | -1.09875100 | 0.86868500 |
| H | 2.41272500 | -0.18014300 | -0.00031700 |
| O | -1.29511600 | -0.67674300 | 0.00000800 |
| Transition State | | | |
| C | -0.55912400 | 0.70526800 | 0.03589600 |
| O | 0.73665500 | 0.66572100 | 0.02770500 |
| H | -0.95052300 | 1.70753000 | -0.10704300 |
| C | -1.23431300 | -0.51930600 | -0.01621500 |
| H | -1.08712800 | -1.19157200 | 0.83110400 |
| H | -0.08612200 | -1.03792700 | -0.47640200 |
| H | -2.25709400 | -0.49171900 | -0.37409400 |
| O | 1.15603100 | -0.67848200 | -0.02666100 |
| VHP | | | |
| C | 0.71121300 | 0.55562700 | 0.01300100 |
| O | -0.64669100 | 0.71005900 | -0.01950400 |
| H | 1.14156500 | 1.55130400 | 0.03470800 |
| C | 1.42255600 | -0.56151700 | 0.00937500 |
| H | 0.96488600 | -1.53787600 | -0.02974100 |
| H | 2.49997000 | -0.48740800 | 0.03579100 |
| O | -1.30543000 | -0.57546700 | -0.09524700 |
| H | -1.79206700 | -0.56741600 | 0.74300300 |

**Table S2.** The absolute energies (a.u.) for each stationary point concerned in the context determined by various methods and the corresponding zero-point energy (ZPE) corrections (a.u.).

| | CH$_3$CHOO | Transition state | VHP |
|---|---|---|---|
| FN DMC | | | |
| E_absolute | -45.727643 | -45.697529 | -45.755426 |
| [a]ZPE | 0.0592218 | 0.0555245 | 0.0595158 |
| ZPE scaling factor | 0.9877[1] | | |
| CCSD(T)//B3LYP/aug-cc-pVTZ | | | |
| E_absolute | -228.5913461 | -228.5613654 | -228.6214679 |
| [b]ZPE | 0.0592592 | 0.0555345 | 0.0594895 |
| ZPE scaling factor | 0.986[2] | | |
| CCSD(T)//B3LYP/6-311+G(2d,p) | | | |
| E_absolute | -228.4813888 | -228.4491055 | -228.5097349 |
| [b]ZPE | 0.0592218 | 0.0555245 | 0.0595158 |
| ZPE scaling factor | 0.9877[1] | | |
| CCSD(T)//M06-2X/aug-cc-pVTZ | | | |
| E_absolute | -228.5906754 | -228.5609204 | -228.6208039 |
| [b]ZPE | 0.0604871 | 0.056512 | 0.0605641 |
| ZPE scaling factor | 0.971[2] | | |
| M06-2X/aug-cc-pVTZ | | | |
| E_absolute | -228.9027284 | -228.8746847 | -228.9377826 |
| ZPE | 0.0604871 | 0.056512 | 0.0605641 |
| ZPE scaling factor | 0.971[2] | | |
| B3LYP /aug-cc-pVTZ | | | |
| E_absolute | -228.9500831 | -228.9216866 | -228.9770797 |
| ZPE | 0.0590381 | 0.0556236 | 0.0592375 |
| ZPE scaling factor | 0.986[2] | | |
| B3LYP/6-311+G(2d,p) | | | |
| E_absolute | -228.9939603 | -228.9639957 | -229.0210755 |
| ZPE | 0.0592218 | 0.0555245 | 0.0595158 |
| ZPE scaling factor | 0.9877[1] | | |
| B3LYP/6-31+G(d,p) | | | |
| E_absolute | -228.9312054 | -228.90087 | -228.9566992 |
| ZPE | 0.0594438 | 0.0557142 | 0.0594981 |
| ZPE scaling factor | 0.9806[3] | | |
| B3LYP/6-31G(d,p) | | | |
| E_absolute | -228.9160912 | -228.8881196 | -228.9422057 |
| ZPE | 0.0595679 | 0.0559793 | 0.0597573 |
| ZPE scaling factor | 0.963[4] | | |
| MP2/6-31G(d,p) | | | |
| E_absolute | -228.2575983 | -228.2313346 | -228.2927851 |
| ZPE | 0.0609094 | 0.0577891 | 0.0607062 |
| ZPE scaling factor | 0.93[4] | | |
| MP2/6-31G+(d,p) | | | |
| E_absolute | -228.2763035 | -228.2482461 | -228.3105047 |
| ZPE | 0.0607356 | 0.0574143 | 0.0603382 |
| ZPE scaling factor | 0.968[2] | | |
| MP2/6-311+G(2d,p) | | | |
| E_absolute | -228.421321 | -228.3958054 | -228.4564057 |
| ZPE | 0.0599065 | 0.0566368 | 0.0596995 |
| ZPE scaling factor | 0.9776 | | |
| MP2/aug-cc-pVTZ | | | |
| E_absolute | -228.5292307 | -228.5054926 | -228.5651145 |
| ZPE | 0.0599214 | 0.0566213 | 0.0597973 |

| | |
|---|---|
| ZPE scaling factor | 0.9792[5] |

a. ZPE orrections for the FN DMC results was determined at B3LYP/6311+G(2d,p)level.

b. ZPE corrections for the CCSD(T) results was determined at corresponding DFT levels used for geometry optimization for the stationary point.

**REFERENCE**

[1] M.P. Andersson and P. Uvdal, J. Phys. Chem. A **109**, 2937 (2005).

[2] I.M. Alecu, J. Zheng, Y. Zhao, and D.G. Truhlar, J. Chem. Theory Comput. **6**, 2872 (2010).

[3] K.T. Kuwata, M.R. Hermes, M.J. Carlson, and C.K. Zogg, J. Phys. Chem. A **114**, 9192 (2010).

[4] R. Gutbrod, E. Kraka, R.N. Schindler, and D. Cremer, J. Am. Chem. Soc. **119**, 7330 (1997).

[5] M.K. Kesharwani, B. Brauer, and J.M.L. Martin, J. Phys. Chem. A **119**, 1701 (2015).